\documentclass[aps,preprint,a4paper]{revtex4-1}
\pdfoutput=1
\usepackage{graphicx}
\usepackage{amsmath}
\usepackage{amssymb}
\usepackage{times}
\usepackage[utf8]{inputenc}
\usepackage{placeins}
\def\nn{\nonumber}
\def\beq{\begin{equation}}
\def\eeq{\end{equation}}
\def\beqna{\begin{eqnarray}}
\def\eeqna{\end{eqnarray}}
\def\bea{\begin{array}}
\def\ea{\end{array}}

\begin{document}
\title{Gain and noise spectral density in an electronic parametric amplifier
with added white noise
}
\author{Adriano A. Batista$^1$ and A. A. Lisboa de Souza$^2$}
\affiliation{
$^1$ Departamento de Física, Universidade Federal de Campina Grande\\
Campina Grande-PB, CEP: 58109-970, Brazil\\
$^2$ Departamento de Engenharia Elétrica, Universidade Federal da Paraíba\\
João Pessoa-PB, CEP: 58.051-970, Brazil}
\date{\today}
\begin{abstract}
In this paper, we discuss the behavior of a linear classical parametric 
amplifier (PA) in the presence of white noise and give theoretical estimates
of the noise spectral density based on approximate Green's functions obtained
by using averaging techniques. 
Furthermore, we give analytical estimates for parametric amplification
bandwidth of the amplifier and for the noisy precursors to instability.
To validate our theory we compare the analytical results with experimental data
obtained in an analog circuit.
We describe the implementation details and the setup used in the  
experimental study of the amplifier. 
Near the threshold to the first parametric
instability, and in degenerate-mode amplification, the PA
achieved very high gains in a very narrow bandwidth centered on its
resonance frequency. 
In quasi-degenerate mode amplification, we
obtained lower values of gain, but with a wider bandwidth that is
tunable. 
The experimental data were accurately described by the predictions of the
model. 
Moreover, we noticed spectral components in
the output signal of the amplifier which are due to noise precursors of
instability. 
The position, width, and magnitude of these components are
in agreement with the noise spectral density obtained by the theory proposed 
here.
\end{abstract}
\maketitle
\section{Introduction}





Parametrically-driven systems and parametric resonance occur in many different
physical systems, ranging from the mechanical domain to the electronic,
microwave, electromechanic, optomechanic, and quantum domains. 
In the mechanical domain we have Faraday waves \cite{faraday1831},
inverted pendulum stabilization, stability of boats, balloons, and parachutes \cite{ruby1996}. 
A comprehensive review of applications in electronics and microwave cavities
spanning from the early twentieth century up to 1960 can be found in
Ref.~\cite{mumford1960}. 
A few relevant recent applications, in micro and nano systems, include
quadrupole ion guides and ion traps \cite{paul90}, linear ion crystals in
linear Paul traps designed as prototype systems for the implementation of
quantum computing \cite{raizen1992ionic, drewsen1998large,
kielpinski2002architecture}, magnetic resonance force microscopy
\cite{dougherty1996detection}, tapping-mode force microscopy
\cite{moreno2006parametric}, axially-loaded microelectromechanical systems
(MEMS) \cite{requa06electromechanical, thomas2013efficient}, torsional MEMS \cite{turner98nature}.
In the quantum domain we could mention  wideband superconducting parametric
amplifiers \cite{eom2012nature}, squeezing
in optomechanical cavities below the zero-point motion \cite{szorkovszky2011},
and parametric amplification in Josephson junctions \cite{castellanos2008amplification}
Parametric pumping has had many applications in the field of MEMS, which have
been used primarily as accelerometers, for measuring small forces and as
ultrasensitive mass detectors since the mid 80's \cite{binnig86}. 
An enhancement to the detection techniques in MEMS was developed by Rugar and
Grütter \cite{rugar91} in the early 90's that uses mechanical parametric
amplification (before transduction) to improve the sensitivity of measurements. 
This amplification method works by driving the parametrically-driven resonator
on the verge of parametric unstable zones. 
 



Here, we study  a classical parametric amplifier both in theory and in 
experiment  with analog electronics.
We investigate signal and idler responses near the onset of the first
parametric instability.
We find the experimental results of gain in the signal and idler responses
accurately described by the theory.
We also investigate the noisy precursors of instability.
We provide an analytical expression for the noise spectral density of
a parametrically-driven oscillator with added noise.
Again, mostly, we have very good agreement between experiment and theoretical
predictions.
The novelty compared with the work of Wiesenfeld et al. in the 80's
\cite{jeffries1985observation, wiesenfeld1985noisy} is that we provide a
simpler theoretical derivation for the noise spectral density (NSD) based on
Green's functions and the averaging method, in which the contribution of both
Floquet multipliers involved are taken into account.
Furthermore, we provide an analytical expression for the noise spectral density
in terms of the parameters of the system and not in terms of the largest
Floquet multiplier, which is left unspecified in their theory.
The analytical calculation of the Floquet multipliers can still be
a daunting problem to overcome.
Another difference, is that below threshold there is no periodic solution,
the system is in quiescent mode, a system that was not treated by Wiesenfeld et al..
Because we add noise to a linear parametrically-driven system below threshold,
we can fit the NSD when there is no pumping with the NSD of a harmonic oscillator with added noise and thus calibrate the noise level.
If we had used Wiesenfeld's theory the measured noise level would be off by 6dB.
In their model there is always a parametric pumping, since the noise is treated
as a perturbation around a deterministic limit cycle of a nonlinear dynamical system.
Hence, the possibility of calibration of the NSD around a harmonic oscillator
limit is not possible, or at least is not clear, in their theory.

%

The contents of this paper are organized as follows. In Sec.~(\ref{theory})
we present our theoretical model, in Sec.~(\ref{setup}) we describe our
experimental setup, in Sec.~(\ref{results}) we present and
discuss our numerical and experimental results, and in Sec.~(\ref{conclusions})
we draw our conclusions.

\section{Theory}
\label{theory}
\begin{figure}[h!]
\centerline{\includegraphics[{scale=0.6}]{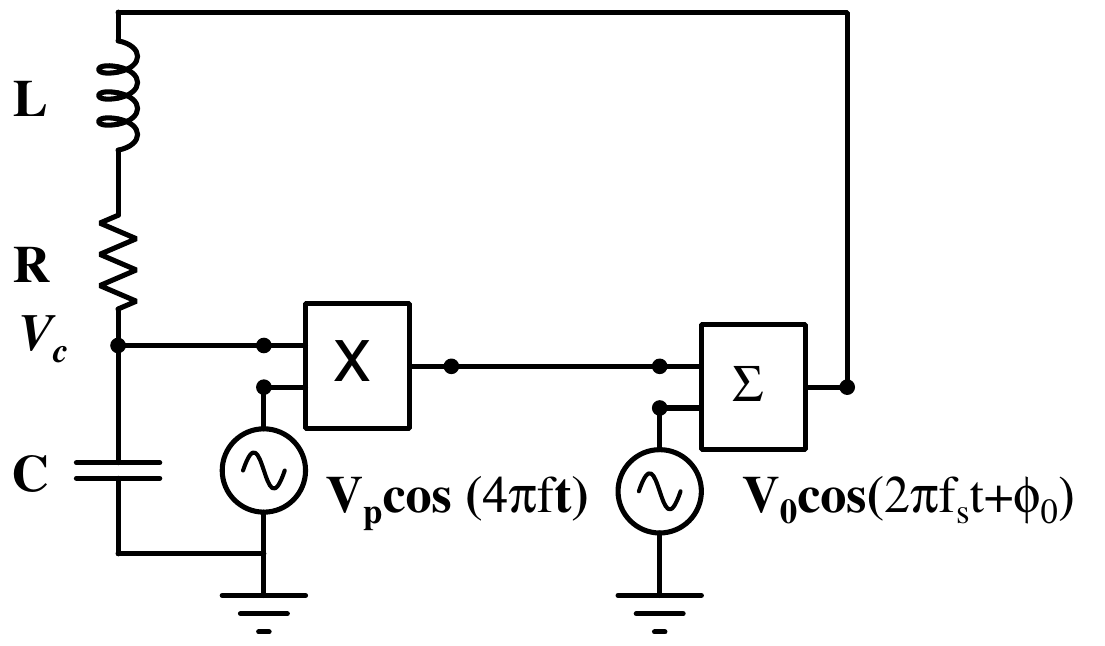}}
\caption{Schematic block diagram of the proposed analog electronic
implementation of the parametric amplifier. The box on the left represents a
multiplier with gain $A$ and the box on the right is a non-inverting summer.}
\label{fig:paramp}
\end{figure}

The  block diagram shown in Fig.~\ref{fig:paramp} is an schematic
implementation of an electronic circuit of the parametric amplifier.
The block with the $\times$ symbol in it represents a multiplier and
the box with a $\Sigma$ symbol is a summer.
The equation that describes this system is given by
\begin{equation}
LC\frac{d^2 V_c}{ds^2}+RC\frac{dV_c}{ds}+V_c= AV_p\cos(4\pi f s)V_c
+V_0 \cos(2\pi f_s s+\phi_0),
\label{paramp_electron}
\end{equation}
where $V_c$ is the voltage on the capacitor, $L$ is the inductance, $C$
is the capacitance, $R$ is the resistance, $V_p$ is the pump amplitude,
$A$ is the amplification factor of the multiplier (with units of
$volt^{-1}$), $V_0$ is the signal voltage amplitude, $2f$ is the pump frequency,
$f_s$ is the
frequency of the signal, and $\phi_0$ is the phase of the
signal.
We can simplify Eq.~\eqref{paramp_electron} with the adimensional time
$t=\omega_0 s$, where $\omega_0=1/\sqrt{LC}$,
$\gamma=1/Q=R/(\omega_0L)$, $F_p=AV_p$, $F_0=V_0$.
We find
\beq
\frac{d^2 x}{dt^2}+x= -\gamma\frac{dx}{dt}+F_p\cos(2\omega t)x+F_0
\cos(\omega_s t+\phi_0),
\label{amp}
\eeq
where $x(t)=V_c(t),\;\omega=2\pi f/\omega_0$, and  $\omega_s=2\pi f_s/\omega_0$.
\subsection{The Green's function of the parametric oscillator}
\noindent
The equation for the Green's function of the parametrically-driven oscillator 
described in Eq.~\eqref{amp} obeys
\begin{equation}
    \left[\frac{\partial^2}{\partial t^2} +1+\gamma \frac{\partial}{\partial t}-F_p\cos(2\omega t)\right]\;G(t,t')=\delta(t-t').
    \label{green_eq}
\end{equation}
We use the retarded Green's function which is $G(t,t')=0$ for $t<t'$.
By integrating the above equation near $t=t'$, 
we obtain the initial conditions when $t=t'+0^+$, which are $G(t,t')=0$ and 
$\frac{\partial}{\partial t}G(t,t')=1.0$.
Assuming the parameters $\gamma, F_p=\mathcal{O}(\epsilon)$, with
$0<\epsilon<<1$, we can write the Green's function in the stable zone of 
the parametrically-driven oscillator in the first-order
averaging approximation \cite{batista2012heating} as
\beq
G(t,t-\tau)\approx G_0(\tau)+G_p(t,\tau),
\label{eq:green}
\eeq
with 
\beqna
G_0(\tau) &=&\frac{e^{-\gamma \tau/2}}{\omega} \left[
\cosh(\kappa\,\tau)\sin(\omega \tau)
+\frac{\delta}{\kappa}\sinh(\kappa\,\tau)\cos(\omega \tau)\right],\nn\\
G_p(t,\tau) &=& -\frac{\beta }{\omega\kappa}e^{-\gamma \tau/2}\sinh(\kappa\,\tau)\cos(\omega (2t-\tau)).\nn
\eeqna
where $\tau=t-t'>0$,
 $\kappa=\sqrt{\beta^2-\delta^2}$, $\beta=-F_p/4\omega$, and 
$\delta=\Omega/2\omega$, and $\Omega=1-\omega^2$.

\subsection{The ac-signal response in the parametric oscillator}
Here we present the theory on classical linear parametric amplification
near the onset of the first instability zone of the parametric
amplifier, i.e. we analyze the response of a parametric oscillator to an
added input ac signal. 
In the following, we will use the Green's function of Eq.~\eqref{eq:green}
to obtain the solution of Eq.~\eqref{amp}.
This is given by
\begin{align}
    x(t)&= \int_{-\infty}^{t}\;G(t,t')F_0\cos(\omega_s t'+\phi_0)\;dt',
\label{ac_response}
\end{align}
since we assume the pump and the input signal have been turned on for a long
time, any homogeneous solution has already decayed.
The signal response is given by
\beqna
&&\int_{-\infty}^{t}dt'\;G_0(t-t')\cos(\omega_s t'+\phi_0) =\mbox{Re}\int_{-\infty}^{t}dt'\;G_0(t-t')e^{-i(\omega_s t'+\phi_0)}\nn\\
&=&\mbox{Re}\left[e^{-i(\omega_s t+\phi_0)}\int_{-\infty}^{t}dt'\;G_0(t-t')e^{i\omega_s (t-t')}\right]\nn\\
&=&\mbox{Re}\left[e^{-i(\omega_s t+\phi_0)}\int_{0}^{\infty}G_0(\tau)e^{i\omega_s\tau}\right]=\mbox{Re}\left[e^{-i(\omega_s t+\phi_0)}\tilde G_0(\omega_s)\right].
\eeqna
In order to obtain
\beqna
\tilde G_0(\omega_s) &=& \int_{-\infty}^\infty e^{i\omega_s t} G_0(t) dt\nn\\
&=&\frac{1}{\omega} \int_{0}^\infty e^{-\gamma t/2}e^{i\omega_s t}\left[
\cosh(\kappa\,t)\sin(\omega t)
+\frac{\delta}{\kappa}\sinh(\kappa\,t)\cos(\omega t)\right]dt,
\eeqna
the following integrals were utilized 
\beqna
\int_{0}^\infty e^{[-\gamma/2+i(\omega_s\pm\omega)]t}\cosh(\kappa t)dt &=&
\frac{\gamma-2i(\omega_s+\omega)}{2[\gamma/2-\kappa-i(\omega_s\pm\omega)][\gamma/2+\kappa-i(\omega_s\pm\omega)]},\nn\\
\int_{0}^\infty e^{[-\gamma/2+i(\omega_s\pm\omega)]t}\sinh(\kappa t)dt &=&
\frac{\kappa}{[\gamma/2-\kappa-i(\omega_s\pm\omega)][\gamma/2+\kappa-i(\omega_s\pm\omega)]},\nn
\eeqna
and we obtained
\beqna
\tilde G_0(\omega_s)&\approx& \frac{1}{\omega}\left\{\frac{1}{2i}\left[
\frac{\gamma-2i(\omega_s+\omega)}{2[\gamma/2-\kappa-i(\omega_s+\omega)][\gamma/2+\kappa-i(\omega_s+\omega)]}\right.\right.\nn\\
&-&\left.\left.\frac{\gamma-2i(\omega_s-\omega)}{2[\gamma/2-\kappa-i(\omega_s-\omega)][\gamma/2+\kappa-i(\omega_s-\omega)]}\right]\right.\nn\\
&+&\left.\frac{\delta}{2}\left[\frac{1}{[\gamma/2-\kappa-i(\omega_s+\omega)][\gamma/2+\kappa-i(\omega_s+\omega)]}\right.\right.\nn\\
&+&\left.\left.\frac{1}{[\gamma/2-\kappa-i(\omega_s-\omega)][\gamma/2+\kappa-i(\omega_s-\omega)]}\right]\right\}.
\label{eq:greenFT}
\eeqna

Hence, with the help of the appendix calculations, we find the stationary 
solution of Eq.~\eqref{amp} to be approximately
\beqna
x(t) &\approx&F_0\left\{\mbox{Re}\left[e^{-i(\omega_s t+\phi_0)}\tilde G_0(\omega_s)\right] 
-\frac{\beta }{2\omega}\mbox{Re}\left[
\frac{ e^{i[(2\omega+\omega_s)t+\phi_0]}}{[\gamma/2-\kappa+ i(\omega+\omega_s)][\gamma/2+\kappa+ i(\omega+\omega_s)]}\right.\right.\nn\\
&+&\left.\left.\frac{ e^{i[(2\omega-\omega_s)t-\phi_0]}}{[\gamma/2-\kappa+ i(\omega-\omega_s)][\gamma/2+\kappa+ i(\omega-\omega_s)]}
\right]\right\}.
\label{eq:x_t}
\eeqna
With $\omega_s\approx \omega$ and when the PA is pumped near the onset 
of the first instability, we find the idler response to be 
\beq
\int_{-\infty}^{t}dt'\;G_p(t, t')\cos(\omega_s t'+\phi_0)\approx
-\frac{\beta }{2\omega}\mbox{Re}\left\{\frac{ e^{i[(2\omega-\omega_s)t-\phi_0]}}{[\gamma/2-\kappa+ i(\omega-\omega_s)][\gamma/2+\kappa+ i(\omega-\omega_s)]}
\right\}.\nn
\eeq
We also find that the envelope of the time series $x(t)$ given in Eq.~\eqref{eq:x_t} is approximately
\beq
\left|e^{-i[(\omega-\omega_s)t-\phi_0]}\tilde G_0^*(\omega_s)-\frac{\beta }{2\omega}\frac{ e^{i[(\omega-\omega_s)t-\phi_0]}}{[\gamma/2-\kappa+ i(\omega-\omega_s)][\gamma/2+\kappa+ i(\omega-\omega_s)]}\right|
\label{eq:envelope}
\eeq

\subsection{Noise spectral density in the parametric oscillator}
We will now investigate the effect of noise on the parametrically-driven 
oscillator \cite{batista2011cooling}.
The parametric oscillator equation with added noise is given by
\begin{equation}
    \ddot x=-x-\gamma\dot x+F_p\cos(2\omega t)\;x+r(t),
    \label{parampNoise}
\end{equation}
where $r(t)$ is a Gaussian white noise that 
satisfies the statistical averages
$\langle r(t)\rangle=0$
and $\langle r(t)r(t')\rangle= 2 D\delta(t-t')$, where $D$ is the noise 
intensity. 
We shall now use the same analytical method developed in Refs.
\cite{batista2011cooling, batista2011signal} to study the parametrically-driven
oscillator with noise as given by Eq.~(\ref{parampNoise}). 

Using the Green's function we obtain the solution $x(t)$ of 
Eq.~(\ref{parampNoise}), which is given by
\begin{align}
x(t)&= \int_{-\infty}^{t}dt'\;G(t,t')r(t'),
\label{x_t}
\end{align}
assuming that the added noise has been turned on for a long
time, hence, any homogeneous solution has already faded out.

The Fourier transform of $x(t)$ is given by
\begin{align}
\tilde x(\nu) &=\int_{-\infty}^\infty e^{i\nu t} x(t) dt=
\int_{-\infty}^\infty dt'r(t')\int_{t'}^\infty dt~ e^{i\nu t}\left[G_0(t- t')
+G_p(t, t')\right]\\
&=\tilde r(\nu)\tilde G_0(\nu)+\int_{-\infty}^\infty dt'r(t')\int_{t'}^\infty dt~ e^{i\nu t}G_p(t, t')\nn
\end{align}

Hence, with the help of the appendix calculations, we find
\beq
\tilde x(\nu) =\tilde r(\nu)\tilde G_0(\nu)-\frac{\beta}{2\omega}
\left[ 
\frac{\tilde r(\nu+2\omega) }{[\rho_-+i(\nu+\omega)][\rho_++i(\nu+\omega)]}+\frac{\tilde r(\nu-2\omega)}{[\rho_-+i(\nu-\omega)][\rho_+ +i(\nu-\omega)]}\right],
\label{noise_response}
\eeq
where $\rho_\pm=-\gamma/2\pm\kappa$ are the Floquet exponents \cite{batista2012heating} in the first-order averaging approximation.
Since the stochastic process of a parametrically-driven oscillator with added
noise as defined in Eq.~\eqref{parampNoise} is cyclo-stationary,
the correlation function of $x(t)$ is not translationally invariant
and the Wiener-Khinchin theorem is not valid.
Therefore, one needs to perform a time average over the usual NSD \cite{wiesenfeld1985noisy}.
One then finds
\begin{align}
    S_x(\nu)&=\overline{\int_{-\infty}^{\infty}d\tau e^{-i\nu \tau}\langle x(t+\tau)x(t)\rangle}
    =\frac{1}{2\pi}\overline{\int_{-\infty}^{\infty}d\nu'e^{i(\nu-\nu') t}\langle \tilde x(-\nu)\tilde x(\nu')\rangle}\nn\\
&=\lim_{\Delta\nu\rightarrow 0^+}\int^{\nu+\Delta\nu}_{\nu-\Delta\nu} 
\dfrac{\langle \tilde x(-\nu)\tilde x(\nu')\rangle}{2\pi}\; d\nu'.
\label{eq:NSD}
\end{align}
With the help of the relation 
$\langle \tilde r(\nu)\tilde r(\nu')\rangle=4\pi D \delta (\nu+\nu')$, 
we find, where $\kappa=\imath\kappa''$ is imaginary 
\beqna
S_x(\nu)&\approx& 2D\left\{|\tilde G_0(\nu)|^2+\frac{\beta^2}{4\omega^2}\left[
\frac{1}{[\gamma^2/4+(\nu+\omega-\kappa'')^2][\gamma^2/4+(\nu+\omega+\kappa'')^2]}\right.\right.\nn\\
&+&\left.\left.\frac{1}{[\gamma^2/4+(\nu-\omega-\kappa'')^2][\gamma^2/4+(\nu-\omega+\kappa'')^2]}\right]\right\}.
\label{eq:noiseSpecDensIm}
\eeqna
or where $\kappa$ is real (nearer the onset of instability),
\beqna
S_x(\nu)&\approx& 2D\left\{|\tilde G_0(\nu)|^2+\frac{\beta^2}{4\omega^2}\left[
\frac{1}{[(\gamma/2-\kappa)^2+(\nu+\omega)^2][(\gamma/2+\kappa)^2+(\nu+\omega)^2]}\right.\right.\nn\\
&+&\left.\left.\frac{1}{[(\gamma/2-\kappa)^2+(\nu-\omega)^2][(\gamma/2+\kappa)^2+(\nu-\omega)^2]}\right]\right\}.
\label{eq:noiseSpecDens}
\eeqna
When the pumping is turned off ($F_p=0$) in Eq.~\eqref{noise_response} and we
take $\omega=1$, we obtain  
\beq
S_x(\nu)\approx\frac{D}{2[\gamma^2/4+(\nu-1)^2]},
\label{eq:NSDharmOsc}
\eeq
which is a very good approximation of the harmonic oscillator noise 
spectral density in high-$Q$ oscillators.
Near the instability threshold and with $\nu\approx\omega$, we obtain
\beqna
\tilde G_0(\nu)&\approx& \frac{1}{4i\omega}
\frac{-\gamma+2\imath[\delta+\nu-\omega]}{[\gamma/2-\kappa-i(\nu-\omega)][\gamma/2+\kappa-i(\nu-\omega)]}
,
\label{eq:greenFunApprox}
\eeqna
and the NSD is given approximately by
\begin{align}
S_x(\nu)&\approx 2D\left\{|\tilde G_0(\nu)|^2
+\frac{\beta^2}{4\omega^2[(\gamma/2-\kappa)^2+(\nu-\omega)^2][(\gamma/2+\kappa)^2+(\nu-\omega)^2]}\right\}\nn\\
&\approx D
\frac{\gamma^2/4+[\delta+\nu-\omega]^2+\beta^2}{2\omega^2[(\gamma/2-\kappa)^2+(\nu-\omega)^2][(\gamma/2+\kappa)^2+(\nu-\omega)^2]}\nn\\
&\approx D
\frac{\gamma^2/4+(\delta+\nu-\omega)^2+\beta^2}{4\omega^2\gamma\kappa}\dfrac{1}{(\gamma/2-\kappa)^2+(\nu-\omega)^2},
\end{align}
which is not exactly a Lorenzian curve as predicted by Wiesenfeld et al.
\section{Apparatus}
\label{setup}

\begin{figure}[!ht]
\centerline{\includegraphics[{scale=0.5}]{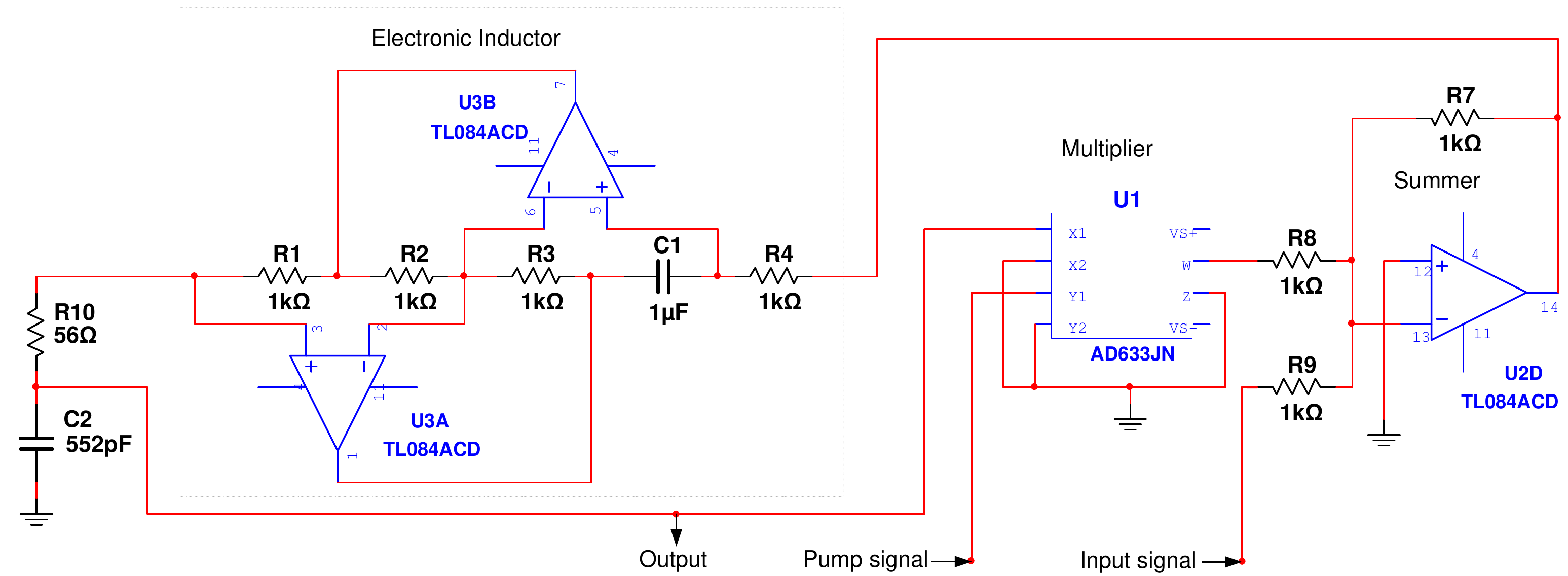}}
\caption{
Electronic circuit implementing the parametric amplifier. 
The 1H electronic inductor is implemented with the aid of operational
amplifiers. 
The IC AD633 represents the multiplier,with a gain of $10V^{-1}$. 
For simplicity, the output buffer and the power supplies are not shown.}
\label{fig:real_amp}
\end{figure}

In this section, we describe the electronic circuit conceived to
implement a parametric amplifier, which is shown in
Fig.~\ref{fig:real_amp}.

\subsection{Analog electronic circuit of the parametric amplifier}

The core of the parametric amplifier is shown in Fig.~\ref{fig:real_amp}
and is comprised of 3 main components: a 4-quadrant analog multiplier (AD633),
which has a conversion gain of $1/10 V^{-1}$, a (weighted inverting) summer
implemented by one operational amplifier (opamp), and an electronic inductor
($1H$) implemented with two opamps (the well-known Antoniou inductor-simulation
circuit).
Unlike the parametric oscillator circuit of Ref. \cite{berthet2002}, we use 
a linear capacitor in place of the nonlinear varicap diode, which makes our
circuit simpler to analyze.
Although nonlinear behavior will eventually appear once the threshold to
instability is crossed, we are interested in the operation of the circuit
as an amplifier in the linear regime, and not as a nonlinear oscillator.

With the choice of $C_2=552$pF, a nominal resonance frequency of $6774Hz$
is obtained. When choosing the capacitor $C_2$, care must be taken to avoid
lowering the quality factor of the resonator (which is otherwise determined
by the quality factor of the inductor along with the value of the
resistor $R10$).
 
The dynamical variable of the circuit is the voltage at the capacitor $C_2$,
whose readout is buffered before being connected to instruments to avoid current
drains during measurements. To avoid outside electromagnetic interference, the 
PA circuit was enclosed in a metallic box.

\subsection{Measurement Setup}

The setup used to characterize the behavior of the parametric amplifier is
shown in Fig.~\ref{fig:setup}.
It is composed of a vector signal analyzer (model $HP89410A$), a waveform
generator (model $33210A$), and an oscilloscope (model $MSO-X 2024A$).
The waveform generator is used to generate the (sinusoidal) pump signal. 
The vector signal analyzer has an internal source that is used as the
signal to be amplified.
Moreover, the analyzer has two input channels, which can be configured to
evaluate, for example, spectrum ratios. 
Besides the frequency-domain analysis  with the $HP89410A$, we have also
observed signals in time-domain with the aid of the oscilloscope.

The setup is flexible enough to enable the experimental characterization of
behavior of the Parametric Amplifier under different operating conditions,
depending upon whether a short-circuit or a signal source is connected to
the inputs of the amplifier.
The different measurement configurations are described below.

\FloatBarrier 
\section{Numerical and experimental results}
\label{results}
\subsection{Harmonic oscillator resonant curve}
To evaluate the response of the harmonic oscillator, for which $F_p=0$ in
Eq.~\eqref{amp}, the pump input port of the amplifier is short-circuited,
while the signal input is fed with a signal source (please refer to
Fig.~\ref{fig:setup}, connections $b$ and $d$). 
We have measured the response (gain) of the harmonic oscillator for a broad 
range of frequencies. 
The results are shown in Fig.~\ref{fig:harmonic_oscillator}.
One can see that the equivalent quality factor Q of the circuit was found 
to be about 65. 
This means that if we consider the circuit of Fig.~\ref{fig:paramp}, the 
equivalent series resistance is about $619\Omega$. 
In addition to the $56\Omega$ shown in Fig.~\ref{fig:real_amp}, there 
are additional losses attributed to the electronic inductor, since the
quality factor of the capacitor $C_2$ was independently measured as about 124. 
The resonance frequency ($f_0=\omega_0/(2\pi)$) was found to be around 
$6400H_Z$, a value less than $6\%$ lower than the nominal frequency.

\subsection{Instability boundary}
After obtaining the resonance curve of the circuit in harmonic oscillator
configuration, we have evaluated the instability boundary of the
circuit in parametric oscillator configuration.
This configuration is obtained by pumping the amplifier while short-circuiting 
its signal input. 
In Fig.~\ref{fig:setup}, this corresponds to connections $a$ and $c$. 
For each value of pump frequency set, the pump amplitude was slowly increased 
from a very low initial value until an oscillation of more than $0.5V$ in 
amplitude was observed at the output of the amplifier.
When comparing the experimental data against the numerical data from 
Eq.~\eqref{amp}, we have found that the equivalent Q of the resonator is 
about 65 (see Fig.~\ref{fig:instability_zone}), a value consistent with the 
one observed for the harmonic oscillator resonant curve.
Moreover, the instability line is centered on the same value of frequency
found for the peak of harmonic oscillator resonant curve (about $6400H_Z$),
which represents half the value of pump frequency for which the lowest pump
amplitude is necessary for instability.

Hereafter the parameter values $Q=\gamma^-1=65$ and $\omega_0\approx 40212$ rad/s will be
used in fitting the experimental data of parametric amplification against
theoretical predictions.

\subsection{Parametric amplification}
Finally, we set the circuit in parametric amplification configuration, by
inputing both pump and signal, as described in Fig.~\ref{fig:setup} with 
connections $a$ and $d$. 
In all the results presented below, the input signal amplitude was limited to 
$2mV$. 
Owing to the high gains which can be obtained, this value has to be kept small
enough to avoid saturation of the amplifier due to limitation of supply
voltage bias and intrinsic nonlinearities of the active components of the
circuit.

\begin{figure}[!ht]
\centerline{\includegraphics[{scale=1.0}]{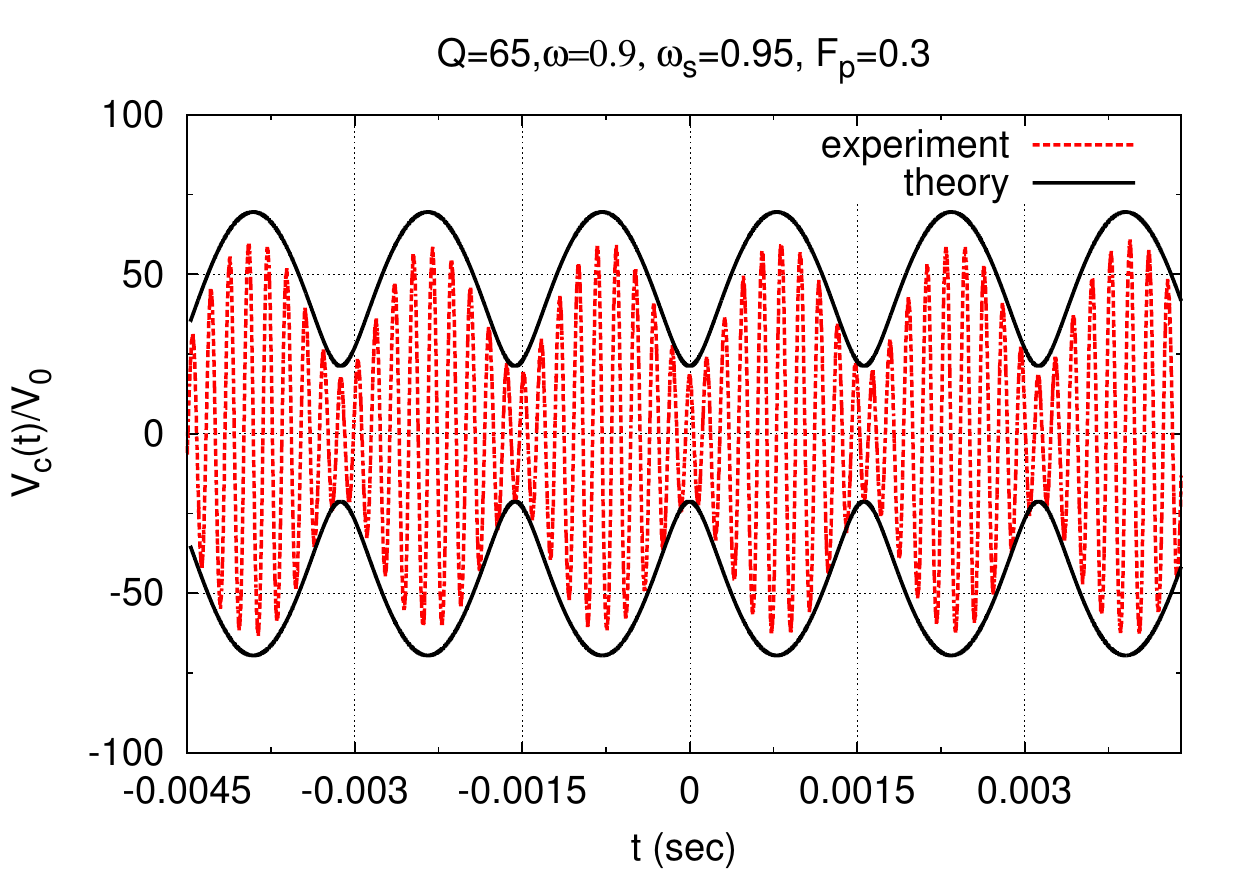}}
\caption{Comparison between a time series from experimental data of $V_c(t)$ and
corresponding envelope obtained from theory in Eq.~(\ref{eq:envelope}). 
The amplification here is set in quasi-degenerate mode with all parameters
indicated in the figure.}
\label{fig:envelope1}
\end{figure}

In Fig.~\ref{fig:envelope1}, we show a time series obtained from the
circuit set with pump amplitude $V_p=3V$, pump frequency $1.8f_0$ and input 
signal frequency $f_s=0.95f_0$, along with the envelope predicted by the 
Eq.~\eqref{eq:envelope} with  $F_p=0.3$.

\begin{figure}[!ht]
\centerline{\includegraphics[{scale=1.0}]{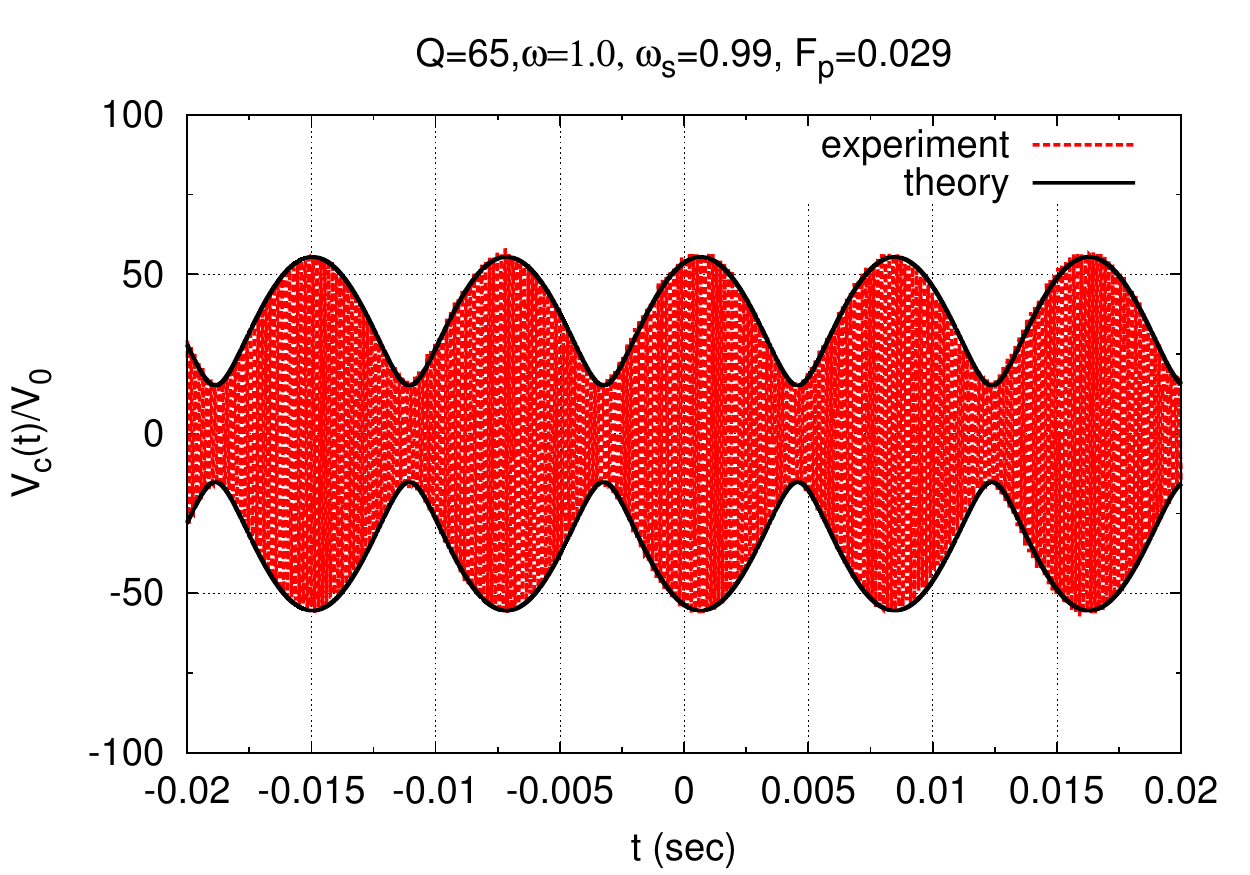}}
\caption{Comparison between time series of experimental data of $V_c(t)$ and
envelope obtained from theory in Eq.~(\ref{eq:envelope}). 
The amplification here is set in degenerate mode with a small detuning between
half the pump frequency, $\omega=1$, and the input signal frequency, $\omega_s=0.99$.
}
\label{fig:envelope2}
\end{figure}

In Fig.~\ref{fig:envelope2}, we show a time series obtained from the circuit 
set with pump amplitude $V_p=0.29V$, pump frequency $2f_0$ and input signal 
frequency $f_s=0.99f_0$, along with the envelope predicted by
the Eq.~\eqref{eq:envelope} with $F_p=0.029$. 
The best agreement between the experimental time series and the theoretical 
envelope is obtained under this condition (degenerate-mode amplification).
This occurs because the accuracy of the perturbative methods used
(averaging and harmonic balance) is higher the smaller the parameter $F_p$ is. 

\begin{figure}[!ht]
\centerline{\includegraphics[{scale=1.0}]{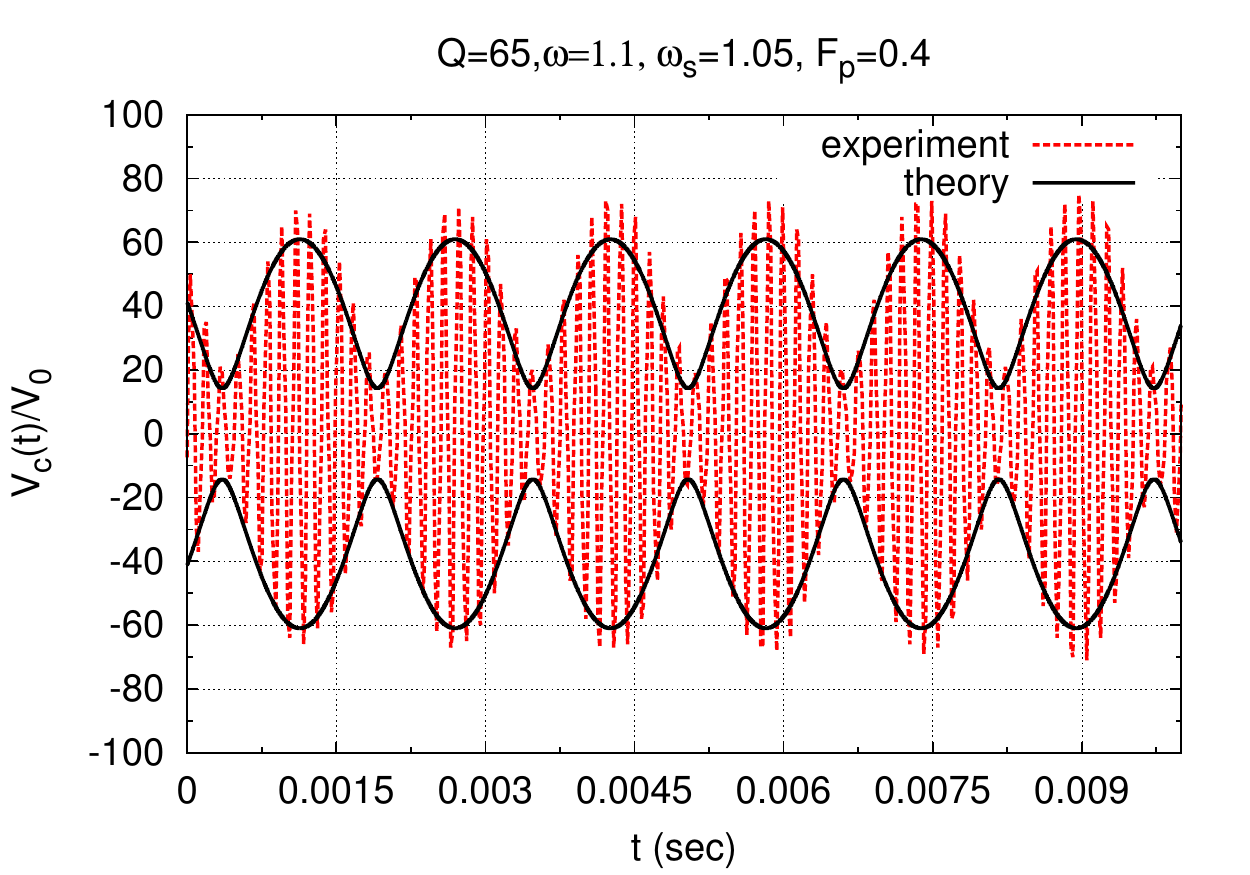}}
\caption{Comparison between a time series from experimental data of $V_c(t)$
and envelope obtained from theory in Eq.~(\ref{eq:envelope}). 
The amplification here is set in quasi-degenerate mode. 
All parameters are indicated in the figure.}
\label{fig:envelope3}
\end{figure}

In Fig.~\ref{fig:envelope3}, we show a time series obtained from the circuit
set with pump amplitude $V_p=4V$, pump frequency $2.2f_0$ and input signal 
frequency $f_s=1.05f_0$, along with the envelope predicted by the 
Eq.~\eqref{eq:envelope} with $F_p=0.4$.

In Fig. \ref{fig:xfft}, we show the signal spectrum measured at the output
of the amplifier with the aid of the signal analyzer (please refer to
Fig.~\ref{fig:setup}). 
The experimental conditions correspond respectively to those associated with
the time series presented in Figs.~\ref{fig:envelope1}, \ref{fig:envelope2},
and \ref{fig:envelope3}.
The signal analyzer computes the spectrum concurrently to the acquisition of
the waveforms by the Oscilloscope.
The experimental results are compared against the numerical Fourier
transforms of the time series $x(t)$ obtained from numerical
integration of Eq.~(\ref{amp}). 
The position and magnitude of the main peaks (signal and idler) is in agreement
with the theoretical predictions of Eq.~\eqref{eq:x_t}.
In frame (b), the middle peak in the experimental data is a noisy
precursor of the parametric instability, whereas in frames (a) and (c)
one hardly notices any effects of noise. 
To help visualize these effects, the insets in frames (a) and (c) show details
of noisy precursors in the experimental data.
Further below we will explain why these precursors are more relevant in the
degenerate-mode amplification of frame (b), than in cases (a) and (c).
In all cases, though, the response of the PA to noise is more pronounced
when the amplifier is tuned to the vicinity of the transition line 
\cite{batista2011cooling, batista2011signal, batista2012heating}. 
Hence, the closer one gets to the transition line, the higher the noisy 
precursor lines in the spectrum will be.
A detailed analysis of the noise spectrum is made in the next subsection.
The noisy precursor effect is in qualitative agreement with the observations of
Jeffries and Wiesenfeld on the effect of broadband noise on the power spectrum
of coherent signals, which was first investigated (theoretically and
experimentally) near period-doubling and Hopf bifurcations in a periodically
driven $p$-$n$ junction \cite{jeffries1985observation, wiesenfeld1985noisy}
and in the context of parametric amplification in Josephson junctions 
\cite{bryant1987nonlinear}.

In Fig.~\ref{fig:signalIdler} one can see the signal and idler responses of the
PA as a function of frequency.
Again, one can see that there is a much broader bandwidth of gain in
quasi-degenerate mode of amplification, such as in frames (a) and (c), than in
degenerate-mode amplification, frame (b).
Moreover, the bandwidth of gain, the peak positions, and the
gain$\times$bandwidth product can be tuned as well, unlike in the
degenerate-mode amplification.
\subsection{Noise spectral density}
\begin{figure}[!ht]
\centerline{\includegraphics[{scale=0.8}]{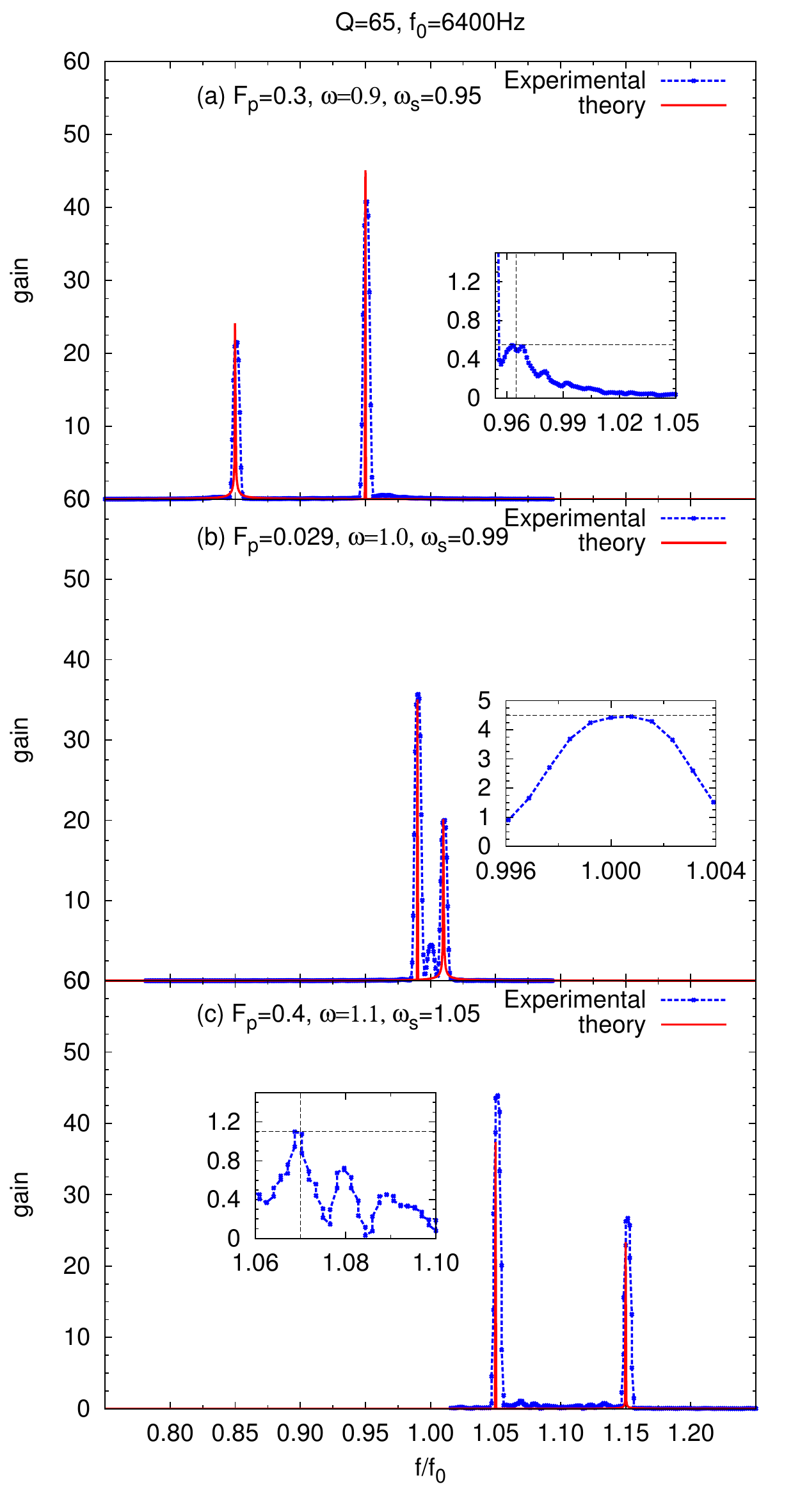}}
\caption{Comparison between experimental data of Fourier transform of
$V_c(t)/V_0$ and numerical Fourier transform of the time series $x(t)/F_0$ from
Eq.~(\ref{amp}).
The Fourier transforms refer to the time series presented in previous figures.
The position and magnitude of the main peaks (signal and idler) is in agreement
with the theoretical prediction of Eq.~\eqref{eq:x_t}. 
The insets in frames (a) and (c) zoom in the noise signatures of the
experimental Fourier transforms.
In frame (b) the middle peak in the experimental data is a noisy precursor of
the parametric instability. 
}
\label{fig:xfft}
\end{figure}

\begin{figure}[!ht]
\centerline{\includegraphics[{scale=0.8}]{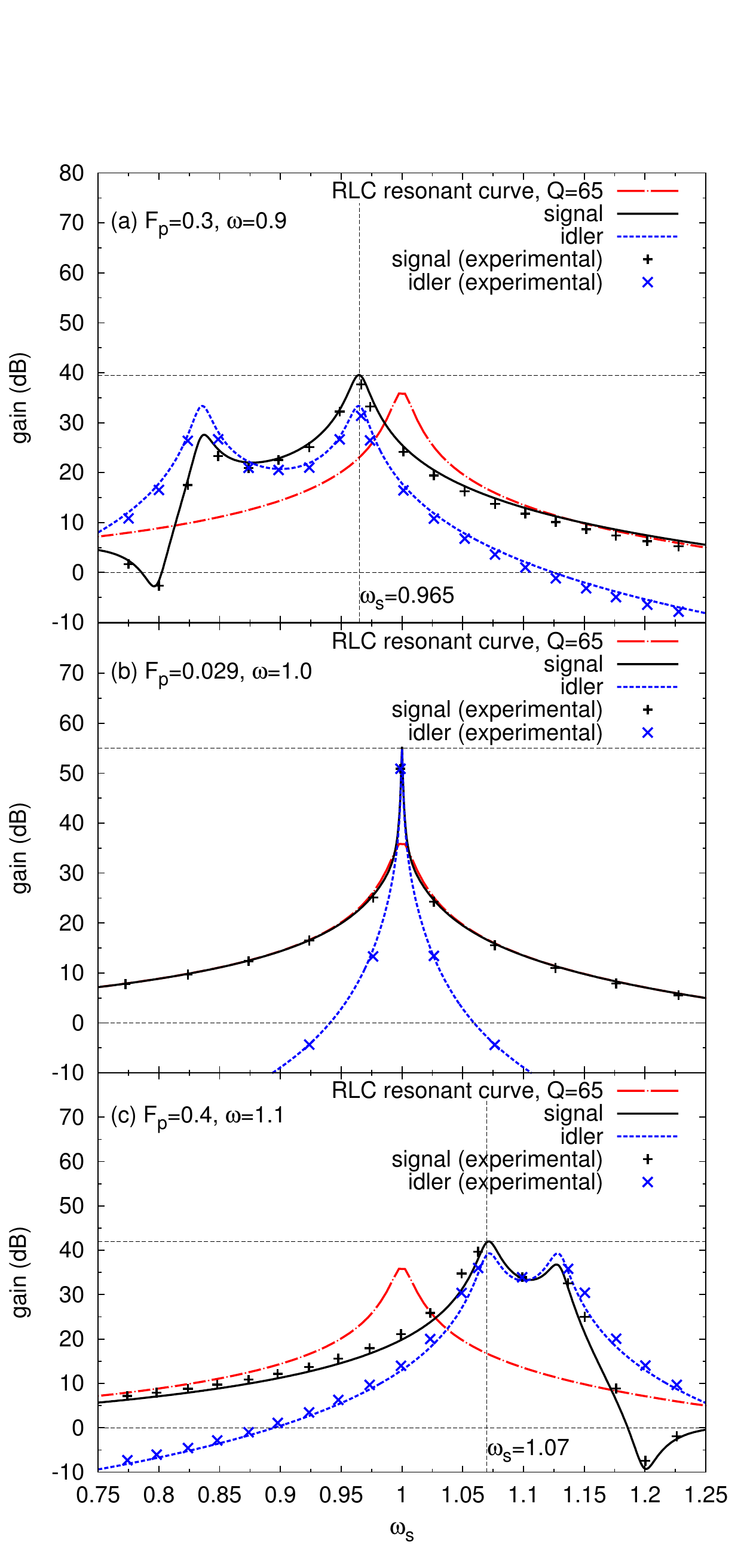}}
\caption{
Theoretical signal and idler response gains compared with RLC resonant curve
gain.
From Eq.~\eqref{eq:x_t} the signal gain in dB scale is given by 
$20\log_{10}(|\tilde G_0(\omega_s)|)$, with $\tilde G_0(\omega_s)$ given in Eq.~\eqref{eq:greenFT}.
Similarly, the  idler gain is obtained
from Eq.~\eqref{eq:x_t}.
In frames a) and c) the Floquet exponent is complex, what causes the split
peaks in the signal and idler responses, whereas in frame b) it is real.
As a comparison, we show the gain of the harmonic oscillator response 
(dot-dashed line).
 }
\label{fig:signalIdler}
\end{figure}

The spectra of the signal responses can be used to predict where
the lines due to noise will turn up in the Fourier transform of the time series
$x(t)$, as shown in Fig.~\ref{fig:xfft}.
Since there is always a small amount of added noise of very broad bandwidth,
which comes along with the input ac signal or is intrinsic to the circuit,
there will be noise components everywhere in the spectra of
Fig.~\ref{fig:signalIdler}.
Hence, as our system is linear, the spectral components of noise will be
amplified in the same way as the input ac signals are amplified.
The strongest case is seen at the peak of degenerate-mode parametric
amplification, when $\omega_s=1$, where the noise will be amplified roughly by
55dB.
The corresponding noise line can be seen in  Fig.~\ref{fig:xfft}(b).
The noise lines in quasi-degenerate-mode parametric amplification are much
smaller, since the the peaks in frames \ref{fig:signalIdler}(a) and (c) have
gains of only roughly 39dB and 42dB, respectively. 
Nonetheless, if one looks in the insets of frames (a) and (c) of
Fig.~\ref{fig:xfft}, one can see elevations in the noise level exactly at
the peaks of the signal response of Fig.~\ref{fig:signalIdler}.
The horizontal and vertical dashed lines at the signal peaks of
Fig.~\ref{fig:signalIdler} are reproduced in the insets of Fig.~\ref{fig:xfft} for clarity.
One can then compare the magnitudes of the noise line peaks of
Fig.~\ref{fig:xfft}.
The difference in gain at noise line peaks between frame \ref{fig:xfft}(b) and
(a) is $20\log_{10}(4.5/0.55)=18.2$dB, whereas  between frame (b) and (c) is
$20\log_{10}(4.5/1.1)=12.2$dB.
On the other hand, the predicted differences in noise peaks based on theory for
the signal gain shown in Fig.~\ref{fig:signalIdler} is 16dB and
13dB, respectively.
These results show that our experimental results are consistent with the
predictions of our proposed linear theory to within an error of under 2.5dB.
We note that the same noise signatures appear in the spectrum when there is
no input ac signal, confirming that the noise precursors are a consequence of
linear response and a consequence of the signal gain spectra of the PA, such
as the ones presented in Fig.~\ref{fig:signalIdler}.

In Fig.~\ref{fig:noiseSpecDens0} we show the noise spectral density for
our circuit setup in harmonic oscillator mode. 
Here we fit the data with a noise intensity of $D= 3.08\times10^{-8}$V$^2$/Hz. 
The source of this noise is intrinsic to the circuit.
Here we showed the exact theoretical result for the NSD
of a harmonic oscillator process given by $S_x^0$ alongside the approximate
result $S_x$ from Eq.~\eqref{eq:NSDharmOsc}.
In our units $S_x$ has dimensions of $V^2$, since the time is adimensional.
In order to set it in proper units, $S_x$ is divided by $\omega_0$.
Both theoretical predictions yield nearly the same spectrum and account well
for the experimental data.
On the other hand, the NSD from Eq. (13) of Ref.~\cite{jeffries1985observation}
gives a result that is 6dB higher than the measured noise. 
Apart from this, their
result can be rescaled to be exactly our result from Eq.~\eqref{eq:NSDharmOsc}
if one divides their prediction by 4.
Here we have a way of measuring the noise level when there is
no pumping, whereas in their model, there is no quiescent solution without noise. 
They developed a perturbative solution due to noise around a limit cycle 
(an isolated periodic solution) and did not calibrate their solution in the
zero pumping limit.

In Fig.~\ref{fig:noiseSpecDens} one can see the comparison between theoretical
predictions, given by Eq.~\eqref{eq:noiseSpecDens} or by Jeffries-Wiesenfeld 
model, and experimental  results for the NSD of a
parametrically-driven oscillator with added white noise.
Here, the noise level used to fit the data was the same one of the harmonic
oscillator configuration.
In frames (a-c) and (e-g) the Floquet exponents are still complex, as can be
seen in the gap between the noise peaks. 
These peaks are located symmetrically with respect to $\omega$ and the distance
between them is twice the imaginary part of the Floquet exponents. 
In frame (d) the Floquet exponents are real and, hence, there is no split peak
in the noise spectrum.
The discrepancy of theory and experiment at the peak of the NSD is certainly
due to nonlinear effects not accounted for by our linear model.
Again here the Jeffries-Wiesenfeld model is off with a level of noise higher
by slightly over 6dB.
Note also that the absence of sharp peaks in the spectrum is because we have a
quiescent solution and not a periodic orbit solution when there is no noise.

\section{Conclusion}
\label{conclusions}
Here we obtained experimental results on  gain and bandwidth of
classical parametric amplification that are quantitatively well
approximated by a theory based on averaging techniques and on
Green's function theory.
Although one can reach extremely high values of gain in degenerate mode
parametric amplification, the corresponding bandwidth is very narrow. 
This is sometimes an undesirable characteristics, since it makes tuning
to the signal a difficult task.
We have found that the bandwidth of the parametric amplifier can be
increased if we set the amplifier in quasi-degenerate-mode
amplification, i.e. with detuning ($\Omega\neq0$).
This comes at the expense of the high gain obtained in  degenerate-mode
amplification.
Guided by the model developed here, the optimal amount of gain and
bandwidth could be found by carefully tuning the pump frequency and the
pump amplitude.
Moreover, the bandwidth of gain, the peak positions, and the
gain$\times$bandwidth product can be tuned as well, unlike in the
degenerate-mode amplification.

Furthermore, we presented a theory for obtaining an
approximate analytical expression for the NSD
of the parametric oscillator with added noise that accounts for the noisy
precursors of instability in the PA.
With the information of the signal and idler gain spectra and the input
noise level, one could, in principle, determine how the noise features
will be manifested in the power spectrum of the output signal of the
PA.
Although Wiesenfeld et al. 
\cite{jeffries1985observation, wiesenfeld1985noisy} developed a general
comprehensive model for accounting for the noisy precursors of bifurcations
of codimension-1 based on Floquet theory, the model we propose is simpler to
apply and has all the theoretical expressions necessary to compare with
experimental results.
It does not depend on generic unspecified Floquet exponents.
It is worthy to mention that the present model describes the simplest
nontrivial system to which the generic theory of Wiesenfeld could be applied
to, but which was actually overlooked in the literature so far to the authors
knowledge.
Furthermore, we have results for the noise spectral density that take into 
account the presence of two Floquet multipliers, and not just one as
described in Ref. \cite{wiesenfeld1985noisy}.
The existence of two complex Floquet multipliers is characterized by the 
appearance of two noise peaks in the NSD spectrum, which can be seen in several of our results.
This is especially relevant in high $Q$ parametric oscillators, where the
stable parameter region in which the Floquet amplifiers are real decreases with
increasing $Q$. 
Hence, it becomes harder to tune into this region, specially, when there is
detuning in the pump frequency with respect to the resonance frequency, so one
has to account for the contribution from both Floquet exponents.
Also, due to this, the noisy precursors spectral curves are not properly
Lorenzian anymore.

The predicted gain curves proposed here can be used to determine signal and
idler gains, the output noise spectral density, and the figure of noise of the
PA (ratio of output and input signal-to-noise ratios of the PA). 
Furthermore, our PA circuit can be used as a simple and inexpensive
experimental platform to test recent theoretical predictions
\cite{batista2011signal, batista2011cooling, batista2012heating} on thermal
noise squeezing.

Finally, the close proximity of the theoretical predictions and the
experimental results obtained here indicates that one could
design electronic devices based on PAs that could achieve extremely
high gains, have very little noise, and be tunable.
Future experimental and theoretical work on the PAs will be performed in
analyzing the effects of nonlinearity on the dynamic range of amplification
and the phenomenon of noise squeezing.
\appendix*
\section{Idler response calculations}
The idler response is given by
\beqna
&&\int_{-\infty}^{t}dt'\;G_p(t, t')\cos(\omega_s t'+\phi_0) 
=\int_{0}^{\infty}d\tau\;G_p(t, t-\tau)\cos(\omega_s (t-\tau)+\phi_0)\nn\\
&=&-\frac{\beta }{\omega\kappa}\int_{0}^{\infty}d\tau\;e^{-\gamma \tau/2}\sinh(\kappa\,\tau)\cos(\omega (2t-\tau))\cos(\omega_s (t-\tau)+\phi_0)\nn\\
&=&-\frac{\beta }{2\omega\kappa}\int_{0}^{\infty}d\tau\;e^{-\gamma \tau/2}\sinh(\kappa\,\tau)\left[\cos\left((\omega+\omega_s)(t-\tau)+\omega t+\phi_0\right)+\cos\left((\omega-\omega_s)(t-\tau)+\omega t-\phi_0\right)\right]\nn\\
&=&-\frac{\beta }{2\omega\kappa}\mbox{Re}\left\{e^{i\omega t}\int_{0}^{\infty}d\tau\;e^{-\gamma \tau/2}\sinh(\kappa\,\tau)\left[e^{i[(\omega+\omega_s)(t-\tau)+\phi_0]}+e^{i[(\omega-\omega_s)(t-\tau)-\phi_0]}\right]\right\}\nn\\
&=&-\frac{\beta }{2\omega}\mbox{Re}\left\{
\frac{ e^{i[(2\omega+\omega_s)t+\phi_0]}}{[\gamma/2-\kappa+ i(\omega+\omega_s)][\gamma/2+\kappa+ i(\omega+\omega_s)]}\right.\nn\\
&+&\left.\frac{ e^{i[(2\omega-\omega_s)t-\phi_0]}}{[\gamma/2-\kappa+ i(\omega-\omega_s)][\gamma/2+\kappa+ i(\omega-\omega_s)]}
\right\},
\eeqna
where we used the following result
\beqna
\int_{0}^{\infty}d\tau\;e^{-[\gamma/2+ i(\omega\pm\omega_s)]\tau}\sinh(\kappa\,\tau)&=&\frac{1}{2}\left[\frac{1}{\gamma/2-\kappa+ i(\omega\pm\omega_s)}-\frac{1}{\gamma/2+\kappa+ i(\omega\pm\omega_s)}\right]\nn\\
&=&\frac{\kappa}{[\gamma/2-\kappa+ i(\omega\pm\omega_s)][\gamma/2+\kappa+ i(\omega\pm\omega_s)]}.
\eeqna
\section{Noise response}
We have to solve
\[
\int_{t'}^\infty dt~ e^{i\nu t}G_p(t, t')\approx 
-\frac{\beta}{\kappa\omega}e^{\gamma t'/2}\int_{t'}^\infty dt~ e^{i\nu t}e^{-\gamma t /2}\sinh(\kappa\,(t-t'))\cos(\omega (t+t')),
\]
\begin{align}
\int_{t'}^\infty dt~ e^{(-\gamma/2+i\nu) t}\sinh(\kappa\,(t-t'))\cos(\omega (t+t'))&=
\frac{e^{i\omega t'}}{2}\int_{t'}^\infty dt~ e^{[-\gamma/2+i(\nu+\omega)]t }\sinh(\kappa\,(t-t'))\nn\\
&+\frac{e^{-i\omega t'}}{2}\int_{t'}^\infty dt~ e^{[-\gamma/2+i(\nu-\omega)] t }\sinh(\kappa\,(t-t'))
\end{align}

\[
\int_{t'}^\infty dt~ e^{[-\gamma/2+i(\omega+\nu)]t }\sinh(\kappa\,(t-t'))=I_1+I_2,
\]
where
\beqna
I_1 &=&\frac{e^{-\kappa t'}}{2}\int_{t'}^\infty dt~ e^{[-\gamma/2+i(\nu+\omega)]t }e^{\kappa\,t}
=\frac{e^{[-\gamma/2+i(\nu+\omega)]t' }}{\gamma-2\kappa-2i(\nu+\omega)}\nn\\
I_2 &=&-\frac{e^{\kappa t'}}{2}\int_{t'}^\infty dt~ e^{[-\gamma/2+i(\nu+\omega)]t }e^{-\kappa\,t}
=-\frac{e^{[-\gamma/2+i(\nu+\omega)]t' }}{\gamma+2\kappa-2i(\nu+\omega)}\nn
\eeqna
\beqna
I_1+I_2 &=& e^{[-\gamma/2+i(\omega+\nu)]t' }\left[\frac{1}{\gamma-2\kappa-2i(\omega+\nu)}-\frac{1}{\gamma+2\kappa-2i(\omega+\nu)}\right]\nn\\
&=& \frac{4\kappa e^{[-\gamma/2+i(\omega+\nu)]t' }}{[\gamma-2\kappa-2i(\omega+\nu)][\gamma+2\kappa-2i(\omega+\nu)]}
\eeqna
\[
\int_{t'}^\infty dt~ e^{[-\gamma/2+i(\nu-\omega)] t }\sinh(\kappa\,(t-t'))=
K_1+K_2,
\]
where
\beqna
K_1 &=&\frac{e^{-\kappa t'}}{2}\int_{t'}^\infty dt~
e^{[-\gamma/2+i(\nu-\omega)]t}e^{\kappa\,t}
=\frac{e^{[-\gamma/2+i(\nu-\omega)]t'}}{\gamma-2\kappa-2i(\nu-\omega)}\nn\\
K_2 &=&-\frac{e^{\kappa t'}}{2}\int_{t'}^\infty dt~ e^{[-\gamma/2+i(\nu-\omega)]t }e^{-\kappa\,t}
=-\frac{e^{[-\gamma/2+i(\nu-\omega)]t' }}{\gamma+2\kappa-2i(\nu-\omega)}\nn
\eeqna
\beqna
\int_{t'}^\infty dt~ e^{i\nu t}G_p(t, t')&\approx& -\frac{\beta}{\kappa\omega}
e^{\gamma t'/2}\left[ (I_1+I_2)e^{i\omega t'}+(K_1+K_2)e^{-i\omega t'}\right]/2\nn\\
&\approx& -\frac{\beta}{2\omega}
\left[ 
\frac{e^{i(\nu+2\omega)t' }}{[\gamma/2-\kappa-i(\nu+\omega)][\gamma/2+\kappa-i(\nu+\omega)]}\right.\nn\\
&&+\left.\frac{e^{i(\nu -2\omega)t' }}{[\gamma/2-\kappa-i(\nu-\omega)][\gamma/2+\kappa-i(\nu-\omega)]}
\right]\nn
\eeqna

\beqna
\int_{-\infty}^\infty dt'r(t')\int_{t'}^\infty dt~ e^{i\nu t}G_p(t, t')
&\approx& -\frac{\beta}{2\omega}
\left[ 
\frac{\tilde r(\nu+2\omega) }{[\gamma/2-\kappa-i(\nu+\omega)][\gamma/2+\kappa-i(\nu+\omega)]}\right.\nn\\
&&+\left.\frac{\tilde r(\nu-2\omega)}{[\gamma/2-\kappa-i(\nu-\omega)][\gamma/2+\kappa-i(\nu-\omega)]}
\right].\nn
\eeqna

%
\newpage
\FloatBarrier

\begin{figure}[!ht]
\centerline{\includegraphics[{scale=0.8}]{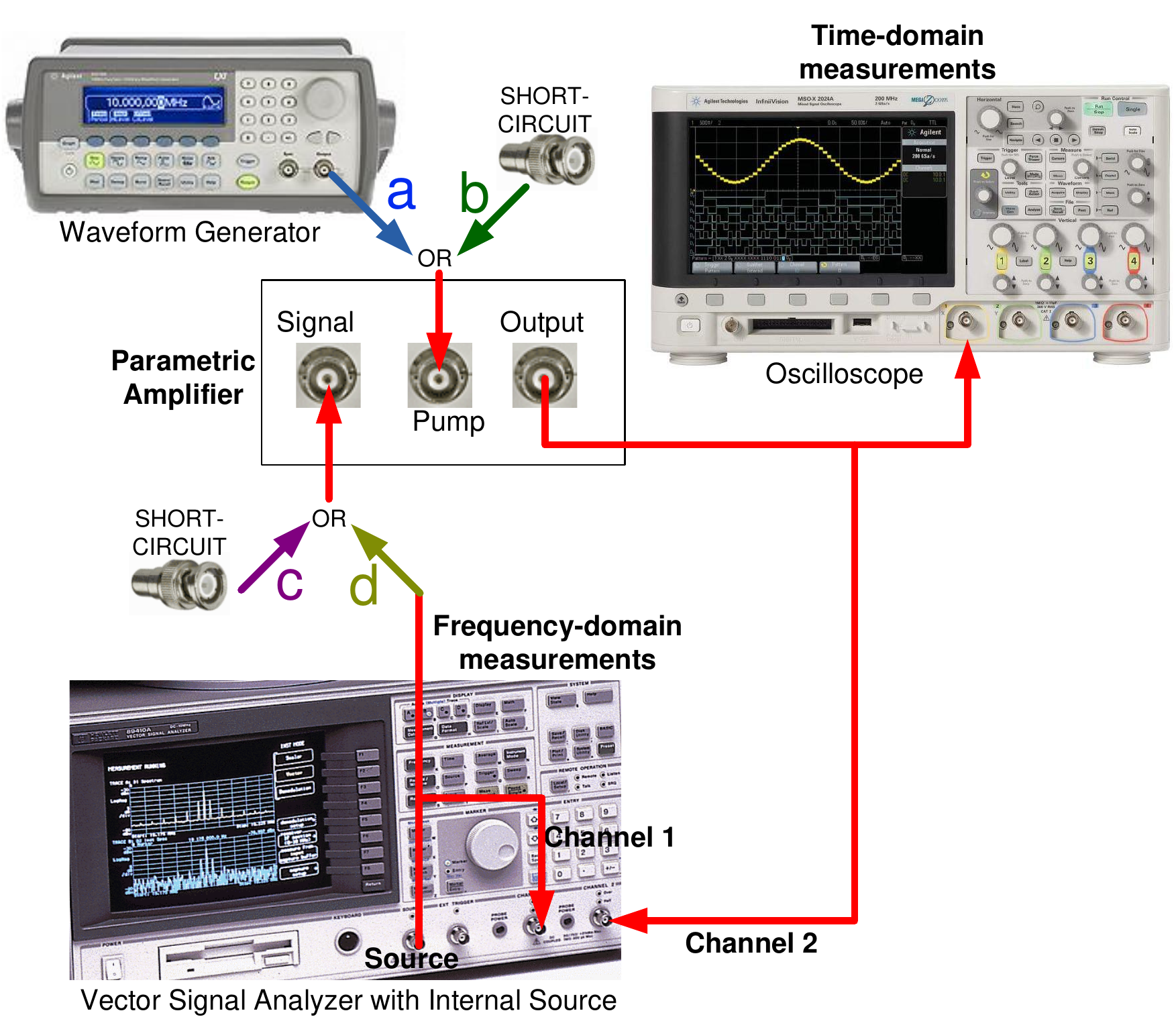}}
\caption{Setup used to characterize the behavior of the parametric amplifier,
allowing different operating conditions. 
With options $b$ and $d$, the PA behaves as a driven harmonic oscillator. 
With options $a$ and $c$, the PA behaves as a parametric oscillator. 
With options $a$ and $d$, the PA behaves as a parametric amplifier if the pump
is kept below the instability threshold.}
\label{fig:setup}
\end{figure}

\begin{figure}[!ht]
\centerline{\includegraphics[{scale=1.0}]{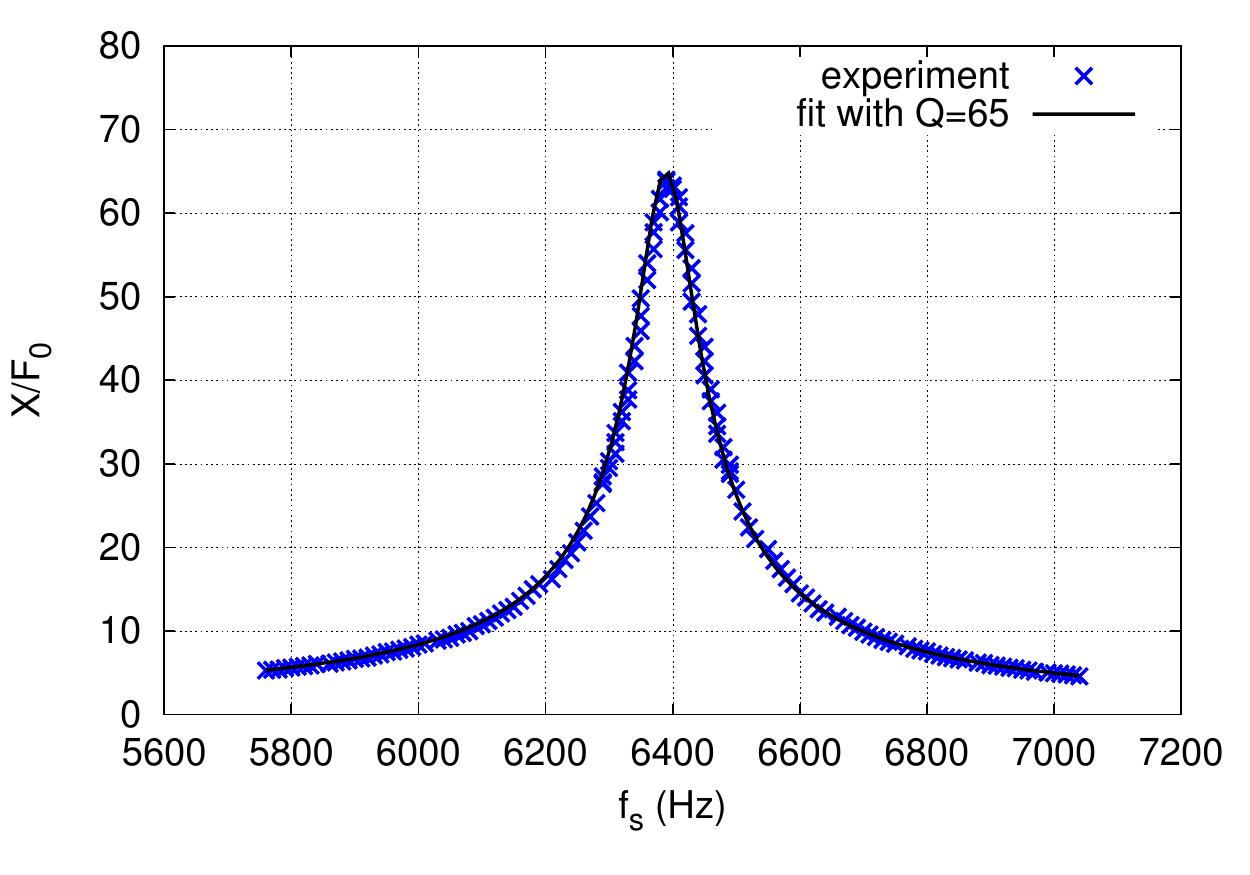}}
\caption{
Harmonic oscillator (pump off) resonant curve, obtained using connections
$b$ and $d$ in Fig.~\ref{fig:setup}.
Here we fit the experimental resonant curve with the theoretic prediction and
find a Q of approximately 65 and a resonant frequency of $f_0=6400$Hz.}
\label{fig:harmonic_oscillator}
\end{figure}
\FloatBarrier

\begin{figure}[!ht]
\centerline{\includegraphics[{scale=1.0}]{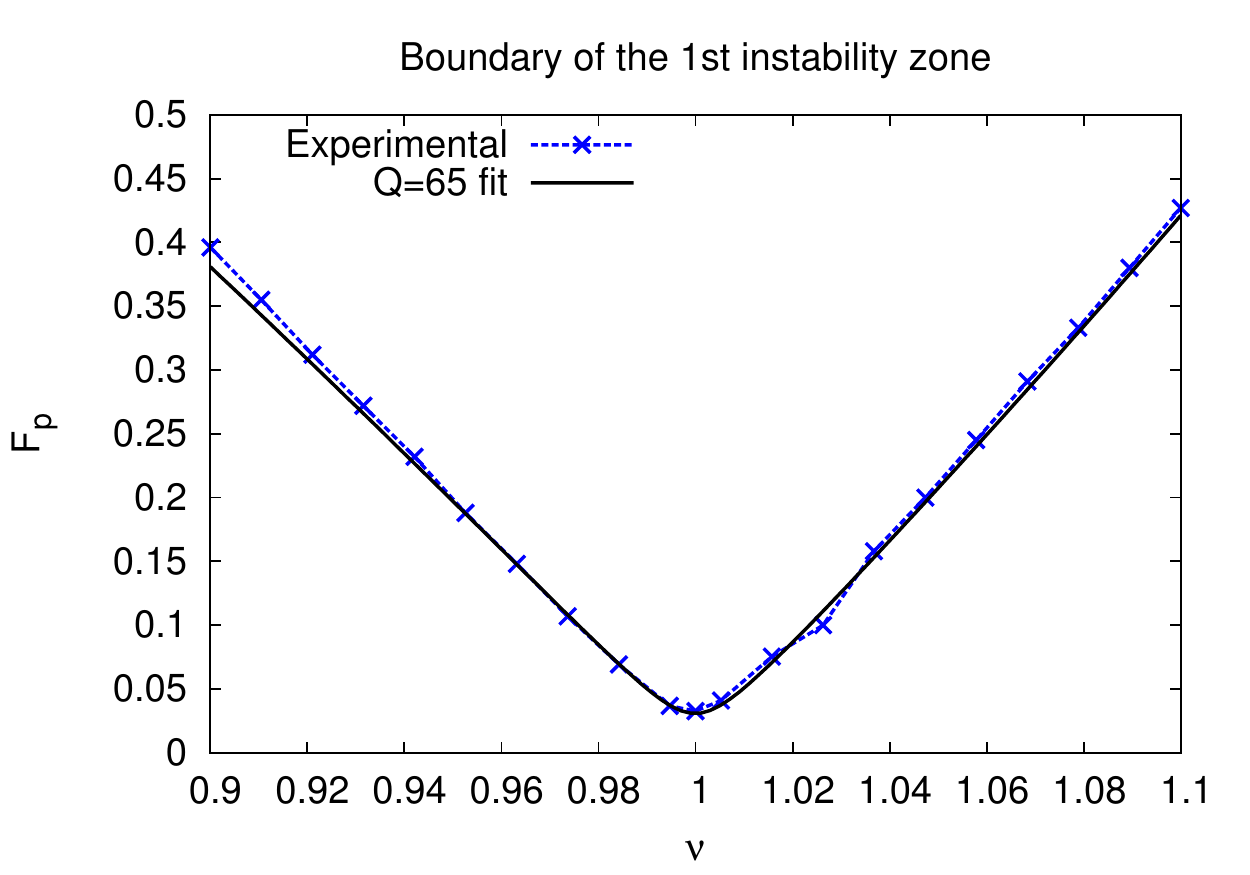}}
\caption{
The instability boundary of the amplifier circuit was obtained with the circuit
set in parametric oscillator mode, as illustrated in Fig.~\ref{fig:setup}.
Above the boundary line, the quiescent state of the circuit ($V_c=0$) is
unstable and the circuit becomes a nonlinear oscillator, whereas below this
line the state $V_c=0$ is stable. 
Very good agreement between experiment and
theory was found with the quality factor $Q=65$.}
\label{fig:instability_zone}
\end{figure}

\begin{figure}[!ht]
\centerline{\includegraphics[{scale=0.8}]{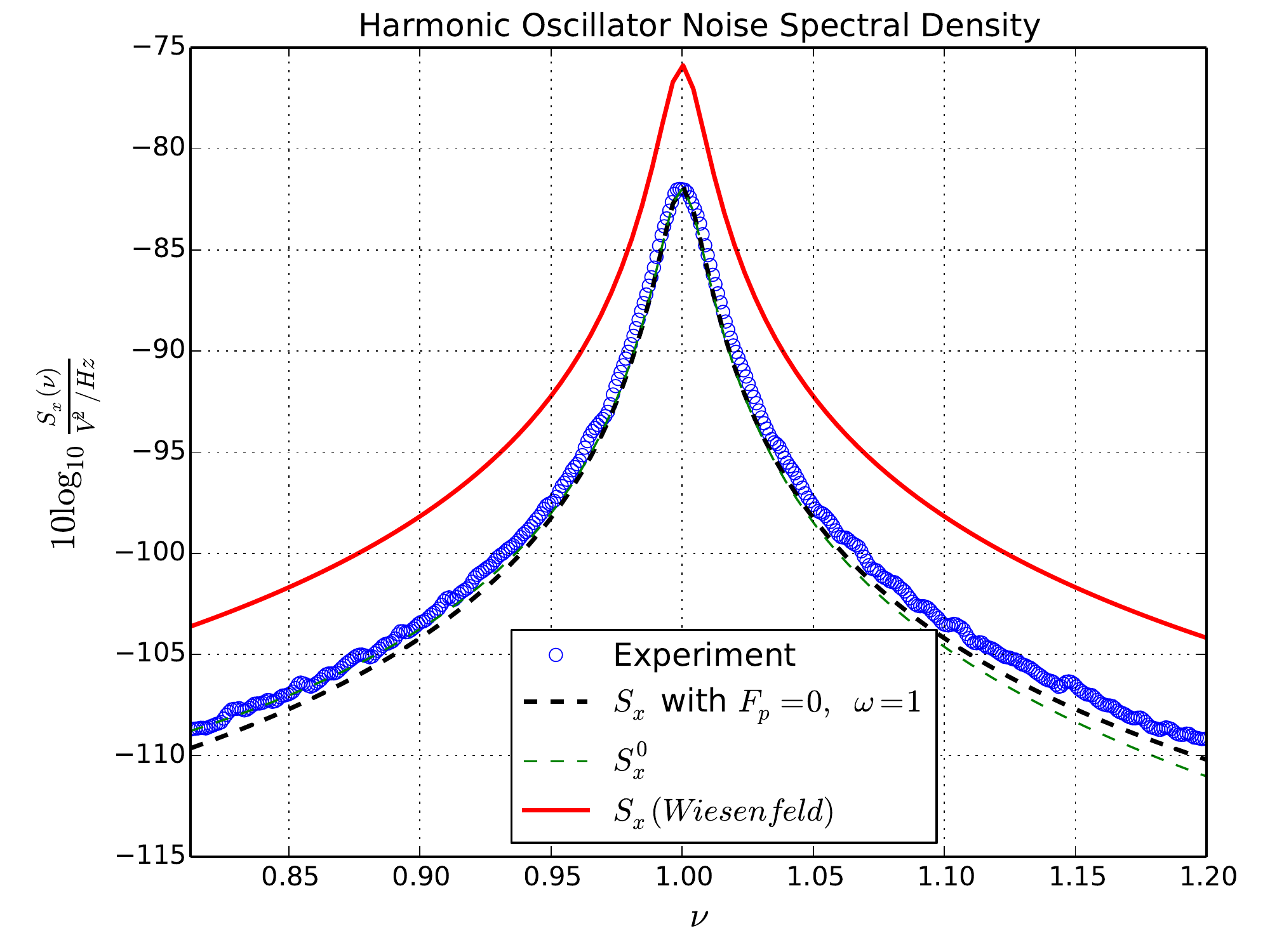}}
\caption{
Comparison between theoretical predictions and experimental  results for the
noise spectral density of a harmonic oscillator.
Here $F_0=0$.
The measuring apparatus depicted in Fig.~\ref{fig:setup} was
setup such that the Signal and the Pump ports were short-circuited 
(options $b$ and $c$).
The results for $S_x$ are given by Eq.~\eqref{eq:noiseSpecDens} without
parametric pumping and at resonance, whereas $S_x^0$ is the exact harmonic
oscillator spectral density.
 }
\label{fig:noiseSpecDens0}
\end{figure}


\begin{figure}[!ht]
    \centerline{\includegraphics[{scale=0.8}]{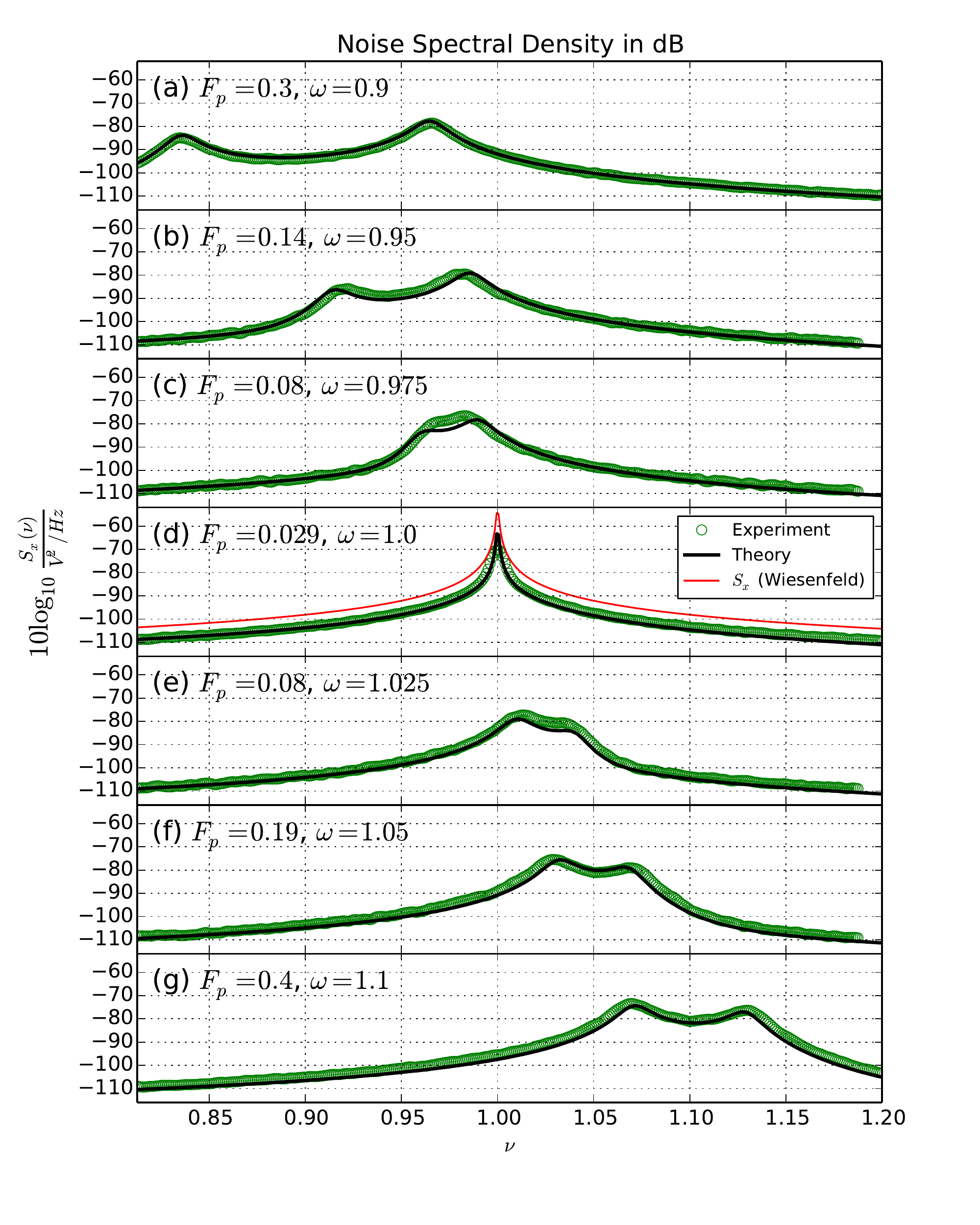}}
\caption{
Comparison between theoretical predictions, given by
Eq.~\eqref{eq:noiseSpecDens} and experimental  results for the noise
spectral density of a parametrically-driven oscillator with added white noise.
Here $F_0=0$.
This was achieved in the setup by short circuiting the input signal (option $c$ in Fig.~\ref{fig:setup}).
The noise level used here is the same one used in the 
harmonic oscillator driven by white noise of Fig.~\ref{fig:noiseSpecDens0}. }
\label{fig:noiseSpecDens}
\end{figure}
\end{document}